\documentclass[pra,aps,twocolumn,showpacs,10pt,floatfix]{revtex4-1}
\usepackage{amsmath,amssymb,graphicx,bm,color,ulem}
\usepackage{amsfonts,amsthm}

\begin{document}

\title{Comment on ``Mode Structure and Orbital Angular Momentum of Spatiotemporal Optical Vortex (STOV) Pulses"}

\author{Miguel A. Porras}
\address{Grupo de Sistemas Complejos, ETSIME, Universidad Politécnica de Madrid, Rios Rosas 21, 28003 Madrid, Spain}

\begin{abstract}
We report a mathematical error and a misinterpretation in https://arxiv.org/abs/2103.03263v4 [Phys. Rev. Lett. {\bf 127}, 193901 (2021)] that has led to a debate about the nature of the transverse orbital angular momentum (OAM) of spatiotemporal optical vortices (STOVs). The transverse OAM of STOVs evaluated theoretically in that Letter is actually only the intrinsic contribution, while the operators used to evaluate the intrinsic and extrinsic contributions are not Hermitian operators as they may lead to complex-valued expectation values.
\end{abstract}

\maketitle

The Letter \cite{MILCHBERG_PRL} commented here raised a debate about the nature and amount of the transverse orbital angular momentum (OAM) of spatiotemporal optical vortices (STOVs) \cite{BLIOKH_PRA_2012,BLIOKH_PRA_2021}. This debate continues today \cite{BLIOKH_PRA_2023,PORRAS_PIERS, HANCOCK_ARXIV}. Angular momentum issues are often very subtle. Seemingly debating about the amount of OAM, the actual substance of the debate, in this author's opinion, is the nature of the transverse OAM. In \cite{MILCHBERG_PRL}, a mathematical error and a misinterpretation led to the incorrect conclusion that STOVs carry a transverse OAM that is purely intrinsic. This error and misinterpretation are also present in the recent preprint \cite{HANCOCK_ARXIV}. A purely intrinsic transverse OAM for STOVs would equate the properties of STOVs with regard to their OAM to those of standard, longitudinal vortices, but such a property is rare and a very peculiar to longitudinal vortices.

By definition, the intrinsic OAM, $L^i$, is the OAM about an axis passing through the object center. For a longitudinal vortex with cylindrical symmetry about the $z$ axis, the fact that its longitudinal OAM is purely intrinsic means that the OAM about any axis parallel to the $z$ axis, $L_z$, takes the same value \cite{BARNETT,BERRY}, i.e., $L_z=L_z^i$. However, for STOVs propagating along the $z$ direction with OAM along a transverse axis, say a $y$ axis, $L_y\neq L_y^i$, with $L_y$ depending on the particular parallel $y$ axis, and implying the existence of an extrinsic transverse OAM, $L_y^e$. This is not just a theoretical consideration, but makes the nature of the OAM of STOVs quite different from that of standard vortices, with implications in how STOVs interact with atoms, particles and light, transmitting or not their OAM.

The transverse OAM operator for the OAM about the transverse $y$ axis in \cite{MILCHBERG_PRL} is $L_y=\xi p_x=-i \xi \partial/\partial x$, where we set $\beta_2=0$ for free space propagation for simplicity, and where $\xi=v_gt-z =ct-z$ in free space. The transverse OAM of an arbitrary envelope field $A(x,\xi)$ is then evaluated as the expectation value
\begin{equation}\label{LY}
\langle L_y \rangle = \int_{-\infty}^{\infty}\int_{-\infty}^{\infty} d\xi dx  A^\star L_y A ,
\end{equation}
where we assume, for conciseness, that $A$ is normalized, $\int_{-\infty}^{\infty}\int_{-\infty}^{\infty} d\xi dx |A|^2=1$. The key point is that $\xi=ct-z$, or better, $-\xi=z-ct$, is not the distance to a fixed transverse $y$ axis, but the distance to an axis moving at the speed of light $c$ accompanying the STOV, i.e., (\ref{LY}) with $L_y=-i \xi \partial/\partial x$ represents the {\it intrinsic} OAM. Indeed (\ref{LY}) coincides with the intrinsic OAM given in \cite{PORRAS_PIERS}.

Next, the operator $L_y$ is written in \cite{MILCHBERG_PRL} in polar coordinates $(\rho,\Phi)$ ($x=\rho\sin \Phi$, $\xi=\rho\cos\Phi$) as
\begin{equation}\label{LYPOLAR}
L_y = -i\left(\rho \sin\Phi\cos\Phi \partial/\partial \rho  + \cos^2\Phi \partial/\partial \Phi \right)\,.
\end{equation}
This is a simple mathematical transformation. There is no physical reason to identify, as in \cite{MILCHBERG_PRL}, the second term with the intrinsic OAM operator, $L_y^i$, and the first term with the extrinsic OAM operator, $L_y^e$. The first term is then said to integrate to zero in the expectation value (\ref{LY}) without specifying any particular form of $A$, concluding that the extrinsic OAM is always zero, and hence that the OAM is purely intrinsic. The first term integrates to zero for the STOV envelope field in Eq. (12) of \cite{MILCHBERG_PRL}, but this only means that the first term in (\ref{LYPOLAR}) does not contribute to the {\it intrinsic} OAM of STOVs.

As demonstrated in the Appendix, the operators $L_y^e= -i\rho \sin\Phi\cos\Phi \partial/\partial \rho$ and $L_y^i=-i\cos^2\Phi \partial/\partial \Phi$ are not Hermitian operators: Their expectation values are generally complex-valued for generic $A(x,\xi)=A(\rho,\Phi)$. Therefore, they cannot represent any physical magnitude. Only the sum of the two operators is Hermitian, the intrinsic transverse OAM. Actually, nothing is said in \cite{MILCHBERG_PRL} about the extrinsic OAM, and therefore about the total OAM.

The total OAM about a fixed $y$ axis passing at an instant of time through the center of the STOV intensity was evaluated in \cite{BLIOKH_PRA_2012}, and has recently been re-evaluated in \cite{BLIOKH_PRA_2023} and \cite{PORRAS_PIERS}. In all cases, it substantially differs from the intrinsic OAM, making STOVs structurally different from their longitudinal counterparts. For completeness, the values of the extrinsic, intrinsic and total OAM of circular STOVs of topological charge $l$ are, in dimensionless units, $-l/2,l,l/2$ in \cite{BLIOKH_PRA_2023}, and $-l/2,l/2,0$ in \cite{PORRAS_PIERS}. The quantitative differences arise only from legitimate, different choices of the STOV center.

Work partially supported by the Spanish Ministry of Science and Innovation under Contract No. PID2021-122711NB-C21.

\section*{Appendix}

The expectation values of $L_y^e = -i\rho \sin\Phi \cos\Phi \partial/\partial\rho$ and $L_y^i = - i \cos^2\Phi \partial/\partial\Phi$ may be complex. Therefore these operators are not Hermitian and cannot represent physical magnitudes.

Using integration by parts in the variable $\rho$ and assuming sufficiently fast decay of $A$ with $\rho$,
\begin{eqnarray*}
\begin{split}
&\langle L_y^e \rangle = -i\int_0^{2\pi} d\Phi \sin\Phi\cos\Phi \int_0^\infty d\rho \rho^2 A^\star \frac{\partial A}{\partial \rho}\\
&                     = i \int_0^{2\pi} d\Phi \sin\Phi\cos\Phi \int_0^\infty d\rho \left(2\rho A^\star + \rho^2 \frac{\partial A^\star}{\partial \rho}\right)A\,.
\end{split}
\end{eqnarray*}
The integral of the second term in the parenthesis can be identified with $\langle L_y^e\rangle^\star$, so we write
\begin{equation}\label{LYe}
\langle L_y^e \rangle = \langle L_y^e \rangle^\star +2i\int_0^{2\pi}d\Phi \sin\Phi\cos\Phi P(\Phi)\,,
\end{equation}
where $P(\Phi) = \int_0^\infty d\rho \rho |A|^2$. The imaginary part of $\langle L_y^e \rangle$ is then
\begin{equation*}
\mbox{Im} \langle L_y^e \rangle = \frac{\langle L_y^e \rangle - \langle L_y^e \rangle^\star}{2i} = \int_0^{2\pi}d\Phi \sin\Phi\cos\Phi P(\Phi)\,.
\end{equation*}
The last integral does not vanish in general. This is even clearer by writing it in Cartesian coordinates:
\begin{equation*}
\int_0^{2\pi}d\Phi \sin\Phi\cos\Phi P(\Phi) = \int_{-\infty}^{\infty}\int_{-\infty}^\infty d\xi dx\, \frac{x\xi}{x^2+\xi^2}|A|^2\,,
\end{equation*}
which does not vanish for $|A|^2$ that presents covariance in the variables $x$ and $\xi$.
Thus, since the expectation value may be complex, $L_y^e$ is not an Hermitian operator.

Using now integration by parts in the variable $\Phi$, and using that $|A(\rho,0)|^2=|A(\rho,2\pi)|^2$,
\begin{eqnarray*}
\begin{split}
&\langle L_y^i \rangle = -i\int_0^{2\pi} \!\!d\Phi \cos^2\Phi \int_0^\infty \!\!d\rho \rho A^\star \frac{\partial A}{\partial \Phi}\\
&                     = i \int_0^{2\pi} \!\!d\Phi \int_0^\infty \!\!d\rho\rho \left(-2\sin\Phi\cos\Phi A^\star + \cos^2\Phi \frac{\partial A^\star}{\partial \Phi}\right)A\,.
\end{split}
\end{eqnarray*}
Again, the integral with of the second term is $\langle L_y^i\rangle^\star$, and then we write
\begin{equation}\label{LYi}
  \langle L_y^i \rangle =  \langle L_y^i \rangle^\star - 2i\int_0^{2\pi}d \Phi \sin\Phi\cos\Phi P(\Phi)\,,
\end{equation}
with $P(\Phi) = \int_0^\infty d\rho \rho |A|^2$ as above. The imaginary part of $\langle L_y^i\rangle$ is then
\begin{equation*}
\mbox{Im} \langle L_y^i \rangle = \frac{\langle L_y^i \rangle - \langle L_y^i \rangle^\star}{2i} = - \int_0^{2\pi}d\Phi \sin\Phi\cos\Phi P(\Phi)\,.
\end{equation*}
which in general does not vanish, as above. Then $\langle L_y^i\rangle^\star$ is not an Hermitian operator either.

Note however that according to (\ref{LYe}) and (\ref{LYi}) the expectation value of $L_y^e + L_y^i=L_y$ is real since the operator $L_y= -i\xi\partial/\partial x$ is Hermitian.

\end{document}